# THE METALLICITY DEPENDENCE OF THE FOURIER COMPONENTS OF RR LYRAE LIGHT CURVES IS THE OOSTERHOFF/ARP/PRESTON PERIOD RATIO EFFECT IN DISGUISE

Allan Sandage

The Observatories of the Carnegie Institution of Washington; 813 Santa Barbara Street, Pasadena, California, 91101, USA

## ABSTRACT

The correlation of particular Fourier components of the light curves of RR Lyrae variables with metallicity, discovered by Simon and later by Kovacs and his coworkers, is shown to have the same explanation as the period ratios (period shifts in log P) between RRab lyrae variables that have the same colors, amplitudes, and light-curve shapes but different metallicites. A purpose of this paper is to demonstrate that the model which predicts the period-metallicity relations in the mediating parameters of colors, amplitudes, and light-curve shapes also explains the Simon/Kovacs et al. correlation between period, $\phi_{31}$, and metallicity. The proof is made by demonstrating that the combination of the first and third phase terms in a Fourier decomposition of RRab light curves, called $\phi_{31}$ by Simon and Lee, varies monotonically across the RR Lyrae instability strip in the same way that amplitude, color, and rise time vary with period within the strip. The premise of the model is that if horizontal branches at the RR Lyrae strip are stacked in luminosity according to the metallicity, then there necessarily must be a log period shift between RR Lyraes with different metallicities at the same $\phi_{31}$ values.

However, there are exceptions to the model. The two metal rich globular clusters NGC 6388 and NGC 6441 with anomalously long periods for their amplitudes, violate the period-metallicity correlations both in amplitudes and in $\phi_{31}$ values (for NGC 6441 where $\phi_{31}$ data exist). The cause must be related to the anomalously bright horizontal branches in these two clusters for their metallicities.

The effect of luminosity evolution away from the age zero horizontal branch, putatively causing noise in the metallicity equations, is discussed. It is clearly seen in the amplitude-period correlations but apparently does not exist in the $\phi_{31}$, period correlation in the data for the globular cluster M3 analyzed by Jurcsik al. and by Cacciari and Fusi Pecci, for reasons not presently understood. Clarification can be expected from study of precision photometric data of evolved RR Lyraes in globular clusters of different metallicity when their Fourier components are known.

KEY WORDS: Stars:Variables:RR Lyrae: evolution: Galaxy: globular clusters-stars: horizontal branch

# I. INTRODUCTION

In an important series of papers, Simon (1988) and independently Kovacs & Zsoldos (1995), Jurcsik & Kovacs (1995, 1996), and Kovacs & Walker (2001) demonstrated that certain combinations of the first few terms of a Fourier series representation of RR Lyrae light curves correlate with metallicity and period. After Simon's 1988 discovery, Kovacs and his co-authors showed that the correlation was continuous with period rather than being simply separated into two period groups, as in half of the original conclusion by Simon. From that they developed the Fourier-component correlation as a new method to estimate RR Lyrae [Fe/H] metallicities using the observational data on periods and particular Fourier components. The evident correlations of certain combinations of Fourier coefficients with [Fe/H] was remarkable, and at first glance, counter-intuitive and mysterious, but now much appears to be transparent.

A purpose of this paper is to show that Simon's (1988, Figs. 1-3) separation of the Fourier data into two period groups, and the development by Kovacs and co-authors into a continuum correlation, is in fact the Oosterhoff/Arp/Preston (sometimes hereafter OAP) period-metallicity effect (Arp 1955; Preston 1959) for cluster and field RR Lyrae variables in disguise, and can be understood by the same model of stacked horizontal branch luminosities for different metallicites that explains the OAP period-metallicity correlation.

The paper is organized as follows. The basic data for the RR Lyraes studied by Simon and Teays (1982) and by Simon (1988) are set out in Table 1 of section 2 on metallicities, periods, amplitudes, rise times, the Fourier combination $\phi_{31}$, and the period shift with metallicity from the $\phi_{31}$-period diagram relative to [Fe/H] = 0.00. Section 3 is a demonstration that $\phi_{31}$ varies monotonically with color and with rise-time across the RR Lyrae instability strip. In section 4 we derive our version of the $\phi_{31}$, log period, metallicity relation where we replace the period in the Kovacs et al. formulation with log P. We contrast it with similar equations that relate amplitude, rise-time, period, and metallicity in the usual OAP formulation of the period-metallicity effect that results from the stacked HB luminosity levels (Sandage 1958, 1981a,b; 1982; 1993a,b, hereafter often S58; S81a,b; S82, S93a,b; Sandage, Katem, & Sandage 1981, hereafter SKS). In these papers we had shown that all the correlations with metallicity can be understood if the luminosity of the horizontal branch is brighter in low metallicity RR Lyrae stars than in the higher metallicity variables, provided that the mediating variables (amplitude, rise time, and now $\phi_{31}$) vary monotonically across the strip.

In Section 5 we point out the caveat raised by the anomalous globular clusters NGC 6388 and NGC 6441 where the periods of their RR Lyraes are long and the amplitudes are large at the observed periods as in the metal poor clusters of the Oosterhoff II group (Pritzl et al. 2000, Fig. 1), yet their metal abundances are high at [Fe/H] ~ -0.5. This anomalous situation shows that exceptions must be expected in the $\phi_{31}$, period, metallicity correlations for these anomalous high metallicity RR Lyraes with long periods and large amplitudes. Such RR Lyrae variables clearly violate, for whatever series of astrophysical reasons (Sweigart and Catelan 1998; Bono et al. 1997 a,b), the $M_V$= f[Fe/H] absolute magnitude-metallicity correlation required by the explanation of the OAP period-metallicity effect using this first approximation posited model.

In Section 6 we show that the effect of evolution away from the age zero horizontal branch (at constant [Fe/H]) should, if the model is correct, introduce scatter in the equations of section 4, imitating a metallicity difference where there is none.

Nine research points discussed in the paper are summarized in section 7.

## 2. THE OBSERVATIONAL DATA FOR 54 FIELD RR LYRAE VARIABLES

Table 1 gives the observational data for the 54 field RR Lyraes for which Simon and Teays (1982) and Simon (1988) have made Fourier decompositions of the light curves and have listed the first several Fourier components from which $\phi_{21}$ and $\phi_{31}$ can be calculated.

Column 1 identifies the particular variable. Column 2 lists the adopted [Fe/H] metallicities determined by Layden (1994), which is on the scale and zero-point of Zinn and West (1984) but with its tie to the field variables improved by Layden from his observations of the K-line intensities.[1] Column 3 is the log of the period. Column 4 gives the amplitude of the V light curve taken from Table 1 of Simon and Teays (1982), and from the listings in the catalog of RR Lyrae properties by Nikolov, Buchantsova, & Frolov (1984, hereafter the Sophia catalog) for eight stars in Simon (1988). Column 5 is the rise-time, in fractional phase units, from minimum to maximum light, read from the light curves plotted in the Sophia catalog. These are similar to, but not identical with, those listed in Table 2 of the Sophia catalog in the "asymmetry" column. Significant differences exist for a few stars where, from the plotted light curves, I could not justify the Sophia listings. Column 6 is the $(B-V)_o$ color, corrected for reddening, based on the observed (B-V) from Table 2 of the Sophia catalog and corrected by E(B-V) calculated from the V absorptions listed by Layden (1994) when divided by the adopted absorption-to-reddening ratio of 3.0. Colors are flagged as more uncertain for those stars where Layden's absorption values are larger than 0.21 mag [i.e. E(B-V) values of 0.07 mag and larger]. Column 7 gives the Fourier combination, $\phi_{31}$, calculated from the relevant Fourier coefficients listed by Simon and Teays (1982) and Simon (1988), defined as

$$\phi_{31} = \phi_3 - 3\phi_1, \qquad (2)$$

by Simon and Lee (1981)[2], where the subscripts denote number of the term in the Fourier expansion. The units are radians. Column 8 is the period shift at constant $\phi_{31}$ relative to [Fe/H] = 0.00 calculated from

$$\Delta \log P = 0.201\phi_{31} - 0.859 - \log P(\text{observed}) \qquad (3)$$

derived later (equation 5 in section 4) relative to a fiducial line for [Fe/H] = 0.00 (Fig. 6 later) for different metallicities.

---

[1] Layden's metallicity scale, used throughout here, differs from the metallicites adopted by Jurcsik and Kovacs (1996, their Table 1) that is based on the scale and zero-point of Blanco (1992) that itself was developed from the scale of Butler (1975). The relation between the Layden K-line metallicity values, [Fe/H]Layden, and those of Jurcsik and Kovacs (1996) is

$$[Fe/H]_{Layden} = 1.05\ ([Fe/H])(J\&K) - 0.20. \qquad (1)$$

Layden's metallicities are more metal poor than those of Jurcsik and Kovacs by about 0.2 dex.

It should also be noted that the Butler/Blanco metallicity scale was used by Sandage (1993a, his Tables 3-6) for field RR Lyraes studied there, but the scale of Zinn and West was used for the RR Lyrae cluster data in the same reference.

[2] To make all phase combinations be in the range of 0 to 2π radians, successive additions of as many 2π values as needed to make φ31 be in the stated range have been made, following the prescription of Simon and Lee.

## 3. SYSTEMATIC VARIATION OF $\phi_{31}$ WITH POSITION IN THE INSTABILITY STRIP

Following the early work of Schaltenbrand and Tammann (1971) in using Fourier series to describe Cepheid light curves, Simon and Lee (1981) introduced the particular combinations of the Fourier coefficients such as $\phi_{31}$ from equation (2) above. Simon (1988), and then Kovacs and co-authors, showed that $\phi_{31}$ was strongly correlated with [Fe/H] at a fixed period. The same is true for $\phi_{21}$, similarly defined, but $\phi_{21}$ is linearly correlated with $\phi_{31}$ which has a larger amplitude in the various correlations. For this reason, $\phi_{31}$ is used exclusively in this paper.

Simon's (1988) demonstration was that $\phi_{21}$ correlates with the Preston (1959) $\Delta S$ metallicity indicator for periods less than 0.575 days, and is separated in the $\phi_{21}$-$\Delta S$ diagram into a discrete period group at high $\Delta S$ for periods greater than 0.575 days.

Simon's division into two period groups with different metallicities is identical with the division of the two Oosterhoff (1939, 1944) period groups of RR Lyrae variables in globular clusters into two metallicity groups by Arp (1955) as confirmed and extended by Kinman (1959).

Preston (1959) then showed that the period-metallicity correlation for field RR Lyraes is continuous, rather than only occurring in two discrete period groups. Later, a continuous variation of period with metallicity in globular clusters, identical to that in the field, was shown to also be followed by the cluster data (Sandage 1993a, Tables 1 to 7 and Figs. 1-9), and more recently by RR Lyrae data in several of the dwarf companions of the galaxy (cf. Siegel and Majewski 2000).

The continuum relations between amplitudes, rise times, and color as they change systematically with period and are shifted in zero-point along the log period axis for different metallicities, have now been known for more than 20 years. These are the period shifts (log P differences) at fixed values of each of the parameters [amplitude, rise times, color--and now $\phi_{31}$] in the standard discussions of the problem (op cit).

The log period shifts are a consequence of the combination of two factors, which are (1) the stacked luminosity levels of the HB depending on metallicity, and (b) the monotonic variation of whichever mediating parameter we choose (amplitude, rise time, and now $\phi_{31}$) with position (i.e color or temperature) in the strip. The explanation of the period shifts follows from the standard model as it is reposited here. The consequence of differences in the luminosity levels is that the periods at the intersection of the lines of constant period, which slant downward in the HR diagram from bright luminosity and blue color to lower luminosities and red color, differ at fixed positions in each strip (i.e. at given values of either amplitudes, color, rise times, and/or $\phi_{31}$) depending on the metallicity (eg. S59; SKS81, Fig. 13). The differences are the period shifts in log P.

The requirement for the model to work is that any particular mediating parameter must vary systematically and monotonically with position within the strip, and be largely independent of [Fe/H] at that position. That this is so for amplitude was found early for the intermediate and metal poor globular cluster M3 (Roberts and Sandage 1955, their Figs. 5-7) compared with M15 (SKS 1981, their Figures 8-12), and then generalized using six clusters (S81a, Table 7); see also Smith (1995).

Proof that a similar monotonic variation of $\phi_{31}$ with position in the strip exists is the subject of this section. Here we show that there is a separation of the $\phi_{31}$-period correlation depending on [Fe/H] (Figs. 3 and 4 later), but that there is no measurable separation with [Fe/H] in the $\phi_{31}$-rise-time and the $\phi_{31}$-color correlations (Fig. 1 and 2 below). These facts are consistent with the model.

We aver that φ31 measures some aspect of light-curve shape. To be sure, more than just the first three terms in a Fourier series expansion are needed to describe all details of the light curves, yet it might be the case that only the first few terms are adequate to describe the bulk of the shape morphology. The rise-time (the degree of asymmetry) is a direct measure of light curve shapes. Small rise times (say 0.1 phase units) go with highly spiked light curves (large asymmetry), and often with large amplitudes. Slow rise times near 0.5 phase units describe symmetrical light curves, and sometimes small amplitude.

Hence, if φ21 and φ31 are also shape parameters, we can expect there to be a correlation between them and rise times, and a general correlation with amplitude, but with larger scatter (amplitude and light-curve shape are not tightly correlated). Because both amplitude and rise-times do vary monotonically with strip position (SKS81, Figs. 8,9,10,and 12) a correlation of amplitude and/or rise-times with the φ31 phase parameter would prove it also varies systematically with position in the strip. Of course, a correlation of φ31 directly with color (which is the position in the strip in the HR diagram) would be the most straightforward proof of a systematic variation of φ31 across the strip. However, color of field RR Lyraes, such as in column 6 of Table 1, carry errors of at least 0.03 mag due to uncertain absorption corrections. Nevertheless, there is a correlation, seen later in Figure 2.

Figure 1 from the data from Table 1 shows the correlation of φ31 with rise time. But because there is such a short range for the variation of rise times between only 0.12 and 0.22 phase units, the observational data must be of exceptional quality to avoid a scatter diagram even if a correlation exists. Hence, the correlation in Fig. 1 is as good as can be expected from the observational errors of the order of 20% (i.e. 0.02 phase units out of only 0.10 phase units for the total range). Nevertheless, the trend is beyond doubt. Small values of φ31 go with highly peaked light curves (small rise times).

Rise times for RR Lyraes in globular clusters are shortest at the blue edge of the strip (SKS 1981, Figs. 6 and 10). Hence, Figure 1 shows that φ31 is smallest at the fundamental blue edge of the strip, increasing in step with higher rise-time numbers (more symmetrical curves) toward the red edge. The four symbols in Figure 1 are for the metallicity ranges listed in the caption. There is no evident separation with [Fe/H].

The more direct proof of a systematic variation of φ31 with position in the strip is in Figure 2 showing the correlation with color. Stars with uncertain colors, so flagged in Table 1, are omitted. The correlation with color is systematic, with a scatter of about 0.03 mag., which is close to the uncertainties of the reddening corrected values. As in Fig. 1, the smallest values of φ31 occur at the blue edge of the strip. There is no evident separation with [Fe/H], again as required by the model.

There is also a correlation of φ31 with amplitude (not shown, but recovered from Table 1), but the scatter is large, betraying a scattered AV-rise-time relation. Nevertheless, smallest φ31 values do occur at the largest amplitudes which are near the blue edge of the strip). Again the scatter is not systematic with [Fe/H].

The conclusion is this: the correlations with rise time, color, and amplitude show that φ31 varies systematically with position in the strip, and is independent of [Fe/H] to first approximation at that position in the strip.

However, when period is used as an independent variable, its correlation with φ31 depends strongly on [Fe/H], which we now show. This is the Simon/Kovacs/Zsoldos/Jurcsik discovery.

## 4. THE CORRELATION OF φ31 WITH PERIOD AND METALLICITY

### 4.1 Three two-parameter correlations at a fixed third parameter

In this subsection we derive the basic equation of the problem that relates the Fourier combination of φ31, log period, and metallicity in a continuum fashion for Bailey type ab RR Lyraes (fundamental mode pulsators). This is the discovery of Kovacs and co-authors. (Note that they use period rather than log period for their formulation, whereas we adopt here a log P dependence that has been standard in earlier discussions of the Oosterhoff effect; eg. S93a).

Consider first the parameter space of φ31 vs. log P in Fig. 3. If the complete data set of Table 1 had been plotted, the face of the diagram would be covered with points in an apparent random scatter. However, the scatter is in fact strongly correlated with [Fe/H], shown by plotting only part of the data in Table 1. In Fig. 3, the data are binned into the two small intervals of [Fe/H]. Individual data points in the intervals of $0.00 > [Fe/H] > -0.60$ (with a mean of -0.35 from 7 stars) and $-1.41 > [Fe/H] > -1.80$ (with a mean of -1.63 from 16 stars) are shown with different symbols. The strong separation of the two metallicity groups is evident. Intermediate values of [Fe/H] give intermediate positions in the diagram. The slope of the lines in Fig. 3 is $\partial \phi_{31}/\partial \log P = 4.970$ at fixed [Fe/H], determined from the adopted three-parameter correlation of equation (4) derived below.

Figure 4 is a different representation of the same data, but now using [Fe/H] as the independent variable. The data are binned into two intervals of log P. Again, intermediate values of log P give intermediate positions in Fig. 4. The 14 stars in Table 1 with log periods between -0.442 and -0.329 (mean = -0.367) are shown by open circles. The 15 stars with log periods between -0.200 and -0.127 (mean = -0.169) are shown as closed dots. The slope of the lines is $\partial \phi_{31}/\partial [Fe/H] = 0.709$ at fixed period, again from equation (4) derived below.

Figures 3 and 4 are not as transparent concerning the Oosterhoff/Arp/Preston effect as the traditional diagram first plotted by Preston (1959) correlating his metallicities of field variables with period. This correlation is shown in Figure 5 for the data in Table 1, similar here to Figures 1 to 4 and 10 of Sandage (1993a) for the RR Lyraes in clusters. The prediction of the model is this: at any given [Fe/H], the correlation of the scatter (i.e. the deviations from some fiducial limit line, shown here by the line in the left panel defining the fundamental blue edge), will be correlated with a parameter that determines the position of a given star in the instability strip. Hence, if the model is correct in its basic premise, the prediction is that there must be a correlation of the horizontal scatter in Fig. 5 (left panel) with amplitude, or color, or rise-time at fixed [Fe/H]. Furthermore, in the present context, if the Simon/Kovacs et al. discovery is to be understood, the scatter must also be correlated with φ31.

The right hand panel shows this is so, seen by dividing the Table 1 data into two small intervals of φ31. There is a clear separation. Again, as in Figs. 3 and 4, if all the data in the left panel of Fig. 5, had been used, the scatter would seem to be a continuous function of φ31. But the point is that φ31 varies systematically across the instability strip as log P changes from the blue edge at short periods, shown by the envelope line in the left panel, to the long periods at the red edge whose envelope line is not shown.

The boundary line in the left panel has been determined from the total data of Layden (1994, his Fig. 1) for 301 field RR Lyraes. The curved relation in Fig. 5 replaces the linear relation of $\log P = -0.117 [Fe/H] - 0.526$ adopted in S93a (Fig. 10 there). It has the equation $\log P = -0.452 + 0.033 ([Fe/H])^2$ derived elsewhere (Sandage 2004).

## 4.2. The adopted three-parameter equation

Using the two-parameter approximations in Figures 3, 4, and 5 at fixed mean values of the third parameter gives, after a least squares iteration,

$$[Fe/H] = 1.411 \phi_{31} - 7.012 \log P - 6.025, \qquad (4)$$
$$\pm 0.014 \quad \pm 0.071 \quad \pm 0.018$$

or

$$\phi_{31} = 0.709 [Fe/H] + 4.970 \log P + 4.270, \qquad (5)$$

determined as follows.

In order to gain a better intuition of the dependencies, we began a step-by-step deconstruction of the scatter, rather than making a one shot blindfold least squares fit in a single global solution for the coefficients of the three variables in equation (4). We first required that both the slopes and the separations of the lines in Figures 3-5 conform to the observed plotted data points in an early version of equation (4).

The separation of the lines in Figures 3 and 4 is determined solely by the coefficient of the log P term in equation (4). Furthermore the slope of the lines in Figure 3 is determined by the ratio of the coefficients of the log P and $\phi_{31}$ terms in equation (4), as is the separation of the lines in Figure 5. Also, the slope of the lines in Figure 4 is determined solely by the reciprocal of the coefficient of the $\phi_{31}$ term in equation (4). With these obvious dependencies, the coefficients of the $\phi_{31}$ and log P terms were first estimated by fits to the slopes and separations of the lines in Figures 3, 4, and 5 using the mean log P, $\phi_{31}$, and [Fe/H] values of the small subsets of the data binned as shown in the diagrams. This gave preliminary values of the coefficients in equation (4). We then used the complete Table 1 data to improve the numerical values by least squares calculations, keeping, in turn, one of the coefficients at its preliminary value and performing the least squares calculation as a linear two-parameter equation on the other two. Iteration gave improved values of each. Two iterations were sufficient to reduce the cross talk between the log P and $\phi_{31}$ terms. That the cross talk is negligible in the final equation (4) is shown by the following tests using the complete data, and dividing the data into two groups of $\phi_{31}$ and then in log P.

Figure 6 shows the test for the slope, $\partial\phi_{31}/\partial\log P$, at constant [Fe/H]. This is the ratio of the coefficients of $\phi_{31}$ and log P in equation (4). The diagram is in the form of the usual formulation of the OAP period shift effect by the dependence of $\phi_{31}$ on log period for various values of [Fe/H]. In accordance with the model (eg. Fig. 3 in S58; Fig. 13 in SKS81), $\phi_{31}$ replaces either amplitude or rise time within the instability strip.

Consider a data point in Fig. 6 with an observed value of [Fe/H] and whose other observed parameters are $\phi_{31}$ and log P. From these data we calculate the difference in log P, shown in Figure 6 as $\Delta \log P$, from the adopted fiducial line, determined from equation (5) to be,

$$\phi_{31} = 4.970 \log P + 4.270, \qquad (6)$$

for [Fe/H] = 0.

Suppose that the slope of the lines of constant [Fe/H] in Figure 6 (and in Fig. 3) is wrong. In that case, the (log P values for given [Fe/H] values will depend on ϕ31. This is the cross talk between log P and ϕ31 at particular values of [Fe/H]. The size of the cross talk will depend on the error in the slope of the ϕ31, log P relation, which is the ratio of the coefficients of the log P and ϕ31 terms in equation (4). That error can be calculated from the test.

Figure 7 shows this test for cross talk. Plotted are the calculated (log P values from equation (6) vs. the observed [Fe/H] values for all stars in Table 1. These (log P values are listed in column 8 of Table 1. Stars with ϕ31 smaller than 2.00 are open circles and those with larger values are skipping-jack crosses. If there is cross talk, the crosses and the open circles will define separate period-shift relations with different slopes.

There is no separation of the open circles and the crosses in Fig. 7 to within the statistics. Neglecting the four stars (AA Aql, VX Her, IU Car, and X Ari) with the largest residuals, (larger than 0.26, all smaller than 0.54 dex), the open-circle data (ϕ31 < 2.00) give the slope of the log P/metallicity correlation as (log P/([Fe/H] = 0.139 + 0.002 (rms = 0.013). For the crosses (ϕ31 > 2.00), neglecting AN Ser, TY Aps, X Crt, TT Lyn, TY Pav, and V445 Oph that have residuals between 0.25 and 0.46, gives the slope to be (log P/([Fe/H] = 0.144 + 0.005 (rms = .017). There is no statistical difference. The same conclusion of no separation is reached by neglecting only the outrageous residual stars of IU Car of the open circles and TY Aps and AN Ser of the crosses, with the results for the slope as 0.145 + 0.002 for the open circles (32 stars) and 0.142 + 0.005 for the crosses (20 stars). Hence, the coefficient of 1.411 for the ϕ31 term in equation (4) needs no correction because the two slope values are the same to within statistics.

A second test for cross talk is to compare the calculated value of [Fe/H] from equation (4) with the observed metallicity values, again separated into two groups by ϕ31. If the calculated minus observed differences in [Fe/H] depend either on ϕ31 and/or log P, then the relevant coefficients in equation (4) will again have cross talk. The test is shown in Figure 8 using log P as the dependent variable.

It there is a variation of ([Fe/H] with log P in Fig. 8, then the coefficient of log P in equation (4) would be incorrect. Alternately, if there is a separation between the open circles and the crosses that varies with log P, then the coefficient of ϕ31 in equation (4) would be incorrect. Neither of these effects are present in Figure 8 to within statistics, seen as follows.

Dividing the data into two groups for log P larger and smaller than -0.3, and combining the open circles and the crosses give nearly identical mean < ([Fe/H]> values for the two period groups. For the short period group, with log P < -0.3 for all ϕ31 values (24 stars) give <([Fe/H]> = +0.017 + 0.037 (rms = 0.178). For the long period group with log P > -0.3, for all ϕ31 (30 stars), <([Fe/H]> = + 0.002 + 0.038 (rms = 0.205). Hence, the slopes to the correlation in Figure 8 for large and small log P values are the same, showing that the coefficient of log P in equation (4) needs no correction.

The second test for a separation between the open circles and the skipping jack crosses in Figure 8 shows <([Fe/H]> = -0.026 + 0.032 (rms = 0.117) over all log P for the 32 stars with ϕ31 < 2.00, and <([Fe/H]> = + 0.059 + 0.044 (rms = 2.02) for the 22 stars with ϕ31 > 2.00 over all log P. If we exclude the 10 most deviant stars (listed earlier in this section), the numbers are <([Fe/H]> = -0.016 + 0.023 (rms = 0.117) for the 28 open circles over all log P, and <([Fe/H]> = 0.000 + 0.025 (rms = 0.097) for the 16 skipping jack stars over all log P. There is no difference in these mean [Fe/H] values. Again the conclusion is that the coefficient of ϕ31 in equation (4) needs no correction.

### 4.3. Comparison with the formulation of Jurcsik and Kovacs

The analysis by Jurcsik and Kovacs (1996) using period rather than log P gave small rms deviations for their sample using their equation (3) of the effect. It is useful to compare their formulation with equation (4) here on the Layden system of metallicities.

We transformed the Jurcisk/Kovacs equation (3) to the Layden metallicity scale by equation (1), and adjusted the constant to give a zero ([Fe/H] mean residual for the calculated minus observed [Fe/H] values. The transformed Jurcisk/Kovacs equation is [Fe/H] = 1.413 (31 - 5.666 P - 5.492 on the Layden metallicity scale. Using all 54 stars in Table 1 over all P in this equation, and subtracting the calculated [Fe/H] values therefrom from the observed values gives <([Fe/H]> = +0.002 + 0.026 (rms = 0.186), which includes all stars no matter how deviant from the mean in Fig. 8. Our value from equation (4) here, using the same stars, gives <([Fe/H]> = + 0.009 + 0.026 (rms = 0.191), nearly identical in the rms accuracies in the two formulations. Again, if we exclude the same ten blatantly deviant stars mentioned earlier, the Jurcsik/Kovacs equation for all (31 over all P, gives ([Fe/H]> = -0.009 + 0.016 (rms = 0.108) for 44 stars, whereas equation (4) for the same sample gives <([Fe/H]> = -0.010 + .017 (rms = 0.109) which again is nearly identical with the Jurcsik/Kovacs formulation. The conclusion is that both formulations give nearly identical results.

### 4.4. Equations similar to equation (4) using amplitude and rise times as the independent parameter

By the arguments in the previous sections it is clear that the Simon/Jurcsik/Kovacs/Zsoldos (31-metallicity correlation is the OAP period shift-metallicity correlation in disguise. Particularly telling is the size of the period shift for given differences in metallicity of (log P/([Fe/H]=0.14 at fixed (31 given by equation (4) and the correlation in Fig. 7. To within the uncertainties, this is the same as 0.12 given by the usual period-shift analysis for the cluster RR Lyraes (S81a, Table 7), and for the field variables (S93a).

It remains only to show that the period-shift formulation in the earlier literature on the OAP effect can also be thrown into the equivalent of equation (4) using amplitude, and/or rise time, and/or color instead of (31.

#### 4.4.1. The period-metallicity correlation using amplitudes

For the demonstration based on the field star amplitudes we have used the data of Layden for the [Fe/H] values and the amplitudes from the Sophia (Nikolov et al. 1984) photoelectric photometric catalog of light curves. We could also have used the period-shift data at constant amplitude for cluster variables from Table 7 of S81b, but the field star data of Layden/Sophia is much more extensive.

A plot of amplitude against log period for the 140 stars in the Layden/Sophia database that have both amplitude and metallicity data shows a scatter diagram (not shown). However, just as in Fig. 3 here, the scatter is a continuous function of [Fe/H]. By again dividing the data into small intervals of [Fe/H] to begin the analysis in a similar way as in Fig. 3, and then as before, making multiple iterations for optimum correlations that give adequately small cross talk, we determine the period/metallicity/amplitude correlation, analogous to equation (4), to be

$$[Fe/H] = -1.453 \, AV - 7.990 \, \log P - 2.145, \qquad (7)$$

+0.027     +0.091        +0.025

where 23 large residual stars with ([Fe/H] > 0.50 are neglected out of a total of 140 stars. The similarity with equation (4) is evident.

The slope of this period-shift/[Fe/H] correlation at constant amplitude is (log P/([Fe/H] = 1/7.990 = 0.125 from equation (7). This is nearly identical with the canonical value of 0.117 derived elsewhere (S93a, Fig. 1 and 10) from the Preston-like log P-metallicity correlation for the field RR Lyraes, showing the ubiquity of the period-shift values using whatever parameter (amplitude, color, rise time, or (31) to measure position in the strip.

Testing equation (7) using the Layden/Sophia database gives an rms scatter of 0.42 dex using all 140 stars. If we exclude the 23 stars with deviations (calculated minus observed) in [Fe/H] larger than +0.50, the rms drops to 0.26 dex for 117 stars. Both are considerably larger than the 0.19 dex and 0.10 dex values for the scatter in Fig. 7 using (31 with and without the large deviation stars. Some of the largest amplitude residuals in the amplitude formulation reach more than 1.00 dex. This is the expected size of the maximum deviation due to evolution from the age zero HB (section 6), but from this field star sample there is no way to flag those stars that are highly evolved from the age zero HB because their individual absolute magnitudes are unknown. Hence, we cannot prove here, as was done elsewhere (S90), that the highly deviant stars (16% of the sample) are deviant because of luminosity evolution.

Although equation (7) is clearly less powerful than equation (4), it is expected to have merit in estimating <[Fe/H]> in individual galaxies using the extensive databases for RR Lyrae variables now becoming available (eg. Siegel & Majewski 2000; Held et al. 2001; Clementini et al. 2003a, b; Mackey and Gilmore 2003).

From the older literature one can recognize that equation (7) describes the separation of the period-amplitude relations of individual families of globular cluster variables by metallicity that became known from the early comparison of the amplitude-period and rise-time period relations in M3 and M15 (SKS81, Figs. 9 and 10) as generalized to six other clusters of varying metallicity (S81b Table 7). The resulting model, the same as posited here, was based on this separation, and could be thrown into to a period-amplitude-absolute magnitude correlation by including the pulsation equation (S81a).

This period-amplitude-metallicity relation has been rediscovered by Alcock et al. (1998) from their gravitational lensing survey near the galactic center. In a second paper, Alcock (2000) used the metallicites determined by Walker and Terndrup (1991) for RR Lyraes in Baade's window to calibrate the evident scatter in their period-amplitude diagram for their large sample of variables (Fig. 7 in Alcock 1998). They derive [Fe/H] = -1.328 AV -8.85 log P - 2.60. Over the relevant period and amplitude range from log P between -0.1 and -0.4 and AV between 0.6 and 1.2 mag, this equation gives results that are generally within less than 0.25 dex deviation from equation (7).

McNamera (1999) used the Alcock et al. 1998 results to show that the RR Lyraes in the Alcock sample exhibit the luminosity-metallicity correlation that is required by the present period-shift model. His slope for the metallicity correlation with luminosities is (MV/([Fe/H] = 0.32 + 0.03, which is within the range of most of the empirical calibrations of this slope by other methods. The power of the McNamera result is its clean proof that metal poor RR Lyraes are more luminous than metal rich variables, as was initially required by the model (S58).

4.4.2. Using rise times

An equation analogous to equations (4) and (7) can also be derived for the rise time as a function of period and metallicity by the same procedure as above. Using the slope of the period-shift, metallicity relation for rise time from Table 7 of S81b, and a linear approximation to the rise time/log P relation in Figure 4 of S81b as RT = 1.44 log P +0.46 for the M3 with [Fe/H] = -1.69, gives, in an obvious way,

$$[Fe/H]= 6.33 \, RT - -9.11 \log P - 4.60. \quad (8)$$

Note that the period shift/metallicity slope at constant rise time is $\partial \log P / \partial [Fe/H] = 1/9.11 = 0.11$, again close to the canonical value of 0.12 derived by other means in S93a.

Although equation (8) is interesting, showing that rise time and amplitude (and indeed color, not discussed here) can each be used to estimate [Fe/H] in a similar way as Simon and Kovacs et al. showed for β31, nevertheless, equation (8) is not expected to be very useful in practice because the range of rise times is so very small for RR Lyraes, and their determination is scarcely better than 20% of the RT value itself, as mentioned earlier.

## 5. THE ANOMALOUS CLUSTERS NGC 6388 AND NGC 6441

Clearly, from the analysis in the previous sections, the posited model here, where the luminosity levels of horizontal branches are stacked for different metallicities, together with the data on the parameters of amplitude, rise times, color, and β31 that each vary monotonically within the strip, explains the features of Figures 1-8 and equations (4), (7) and (8). The lines of constant period slope downward from the upper left to lower right in the HR diagram, cutting the instability strips at different places, and therefore at different parameter values for given periods, depending on the luminosity level of the branches.

The requirement for the model to work is that the luminosity of the age zero horizontal branches must be monotonically brighter with decreased metallicity. Until recently, all observational data (such as the period-amplitude and period-color relations being unique functions of metallicity) satisfied this requirement.

However, two clusters were discovered in 1997 to violate the correlations. RR Lyrae stars in the metal rich ([Fe/H] ~ -0.5) clusters NGC 6388 and NGC 6441 have abnormally long periods for their amplitudes (Pritzl et al. 2000, 2001), and both have abnormally blue extended horizontal branches never before seen in other high metallicity clusters (Rich et al. 1997). From the period-amplitude relations of their RR Lyraes, it is evident that their horizontal branches must be considerably more luminous than other clusters of similarly high metallicites, violating the canonical luminosity/metallicity relation known for all other clusters. Clearly, the RR Lyraes in these two clusters will violate equation (7) for the period, amplitude, metallicity correlation. It is also evident from the β31 data in NGC 6441 by Pritzl et al. (2001) that equation (4) is also blatantly violated, seen as as follows.

Tables 3 and 6 by Pritzl et al. (2001) give AV for 18 and β31 for 11 of the numerous RRab Lyraes in NGC 6441. The amplitude equation (7) predicts a cluster metallicity of <[Fe/H]> = -2.10 + 0.08 (rms = 0.321) from the tabulated data, whereas the observed metaliicity is <[Fe/H]> = - 0.53 + 0.11 (Armandroff & Zinn 1988) on the scale of Zinn and West; a clear contradiction. Equation (4) using the β31 values for 11 variables predicts <Fe/H> = -1.17 + 0.04 (rms = 0.125), which again is a clear contradiction of equation (4) compared with the observed metallicity.

The proximate reason is that the luminosity level of the NGC 6441 HB must be ~ 0.25 more luminous than the "normal" (i.e for the totality of clusters known to date, except also for NGC 6388) high metallicity Galactic globular clusters. The reasons remain obscure, but it is evident that there must be special evolutionary channels causing the abnormally bright horizontal branches for such metal rich clusters. The problem is not presently understood at a fundamental level of the stellar interior physics, but it is under discussion (cf. Sweigart & Catelan 1998, Bono et al. 1997a,b) on several fronts.

There is, then, the caveat that there is more to the complete story of a $M_V$ = f([Fe/H]) relation for RRab Lyrae stars than is given by equations (4) and (7) and by all known clusters except NGC 6388 and NGC 6441. and equations (4) and (7). That such evidently bright RR Lyraes with high metal abundance and long periods for their amplitudes are rare is seen from the fact that few if any are presently known in the field.

## 6. THE EFFECT OF EVOLUTION OFF THE AGE ZERO HORIZONTAL BRANCH

### 6.1. Effect in the amplitude formulation in equation (7)

Proof that the horizontal branches of globular clusters have a true intrinsic width in luminosity due to evolution was made elsewhere (Sandage 1990) by showing that the apparent magnitude spread from the lower envelope of the HB CMD, brightward, is correlated with the increased period of the same stars in the period amplitude diagram, at fixed amplitude. This observed increased luminosity, coupled with increased period at a given place in the instability strip (i.e at a fixed amplitude) was shown to be at the expected level as predicted from the pulsation equation.

Clearly, the deviation of a star in the period-amplitude relation due to evolution at fixed [Fe/H] will introduce noise in the [Fe/H], AV, log P relation of equation (7). In a given cluster with fixed [Fe/H], if log P is shifted from its age zero HB main correlation line in the period-amplitude relation at fixed AV, then the predicted [Fe/H] from equation (7) at fixed AV will be incorrect. This is the noise in the equation due to evolution.

An estimate of the severity of the noise is this; typical observed maximum log period shifts from fiducial lines in period-amplitude diagrams in a variety of clusters (cf. S90) are (log P = 0.10 (S90 see Figs. 9 and 19 for M3, Figs. 10 and 18 for M15, Figs 11 for NGC 6981, Fig. 12 for NGC 6171, and Fig. 13 for M4, all in S90). Therefore, for a log period shift of 0.10 dex at fixed AV, equation (7) predicts an error of 0.8 dex in the predicted [Fe/H] value, the calculated value being more metal poor than the real (observed) value. This is the same order as the observed maximum deviations of the stars in the Layden field sample used in section 4.4.1 used in the derivation of equation (7). Indeed, the maximum resdiuals reach 1.2 dex in the calculated minus observed [Fe/H] residuals, although there are only a few larger than 0.6 dex, corresponding to the modest log period shift of 0.07 dex which is close to the mean of the observed variation of log P from the mean in the cluster data just cited.

However, there is a problem with this explanation. All log period shifts due to luminosity evolution are expected to be positive (observed minus the fiducial period along the unevolved line in the amplitude-period diagrams for age zero HB stars) for the stars that deviate in the period-amplitude diagrams. Such positive log P residuals give only negative residuals in equation (7), whereas there are as many positive as negative residuals in the field star ([Fe/H] data in section 4.4.1. Hence, the positive [Fe/H] residuals must have another explanation, perhaps due to an intrinsic scatter in the age zero HB period-amplitude diagram. In support, we note that this intrinsic scatter is of the order of (log P = + 0.03 dex, giving an intrinsic scatter even for zero evolution, of plus/minus ([Fe/H] =0.2 dex from equation (7). Evolution makes this intrinsic noise larger by its effect, but always in the negative direction for increased luminosity due to evolution. In any case, there is no question that luminosity evolution produces longer periods at given amplitudes, seen in all period/amplitude diagrams for clusters, for constant [Fe/H] and therefore produces noise in equation (7) at fixed [Fe/H].

## 6.2 Expected effect in the equation (4) (31 formulation

We fully expected there would be similar noise due to luminosity evolution as it impacts the (31, period, metallicity relation of equation (4). Indeed, if (31 is a unique function of color in the strip, regardless of metallicity and of luminosity, (assumptions of the model and supported to within the considerable scatter in Figs. 1 and 2), then for stars evolved off the AZHB, compared with stars on the AZHB of the same strip position (and hence the same (31 by the assumptions), their periods should be longer from the pulsation equation for the same reasons as in the amplitude case, and hence the calculated [Fe/H] from equation (4) should be incorrect, again causing noise due to evolution.

However, we were astounded during a rewriting of a late draft of this paper, by the result of Cacciari (private communication) of its evident absence in the M3 RR Lyrae data analyzed by Jurcsik et al. (2003), and now by Cacciari and Fusi Pecci. There is no period shift in a log period/(31 diagram as a function of the deviation of the luminosity due to evolution from the unevolved AZHB. This means that some assumption made in the present model, such as an independence of (31 values with strip position on metallicity and luminosity (the Fig. 1 and 2 proofs to within their considerable scatter), must be in error.

Resolution of this unanticipated problem can be expected from study of the developing literature on precision photometry of RR Lyraes in a variety of globular clusters of different metallicities and different luminosity levels of their evolved stars from the AZHB. Is there a different variation of (31 with color (strip position) for AZHB stars and their evolved daughters? In different clusters of different metallicities is there a different (31/color relation along the age zero HB that depends on metallicity (such as appears to be absent in Fig. 3 to first order)? These and similar questions should have observational solutions when the Fourier, color, and magnitude data eventually are known with high accuracy in many clusters. A purpose of this paper is to point to the problems, not to their solution which presumably will improve the present model.

## 7. SUMMARY

There are nine principal research points in this paper.

(1). The combination of the first and third phase terms in a Fourier series representation of RR Lyrae light curves, defined by Simon and Lee (1981) as (31, varies monotonically across the instability strip, being smallest at the blue fundamental edge and increasing toward the red edge (Figs. 1 and 2).

(2). Empirically, φ31 also varies systematically across the scatter in the log P, [Fe/H] correlation (Fig. 5), showing that this scatter is largely suppressed by using φ31 as a correlating parameter. This is the Simon/Kovacs/Zsoldos/Jurscik discovery, but described in the language of the Oosterhoff/Arp/Preston period-metallicity correlation that uses amplitude, or rise time, or color (S81a,b; 93a) as the mediating variable instead of φ31.

(3). The difference between [Fe/H](calculated) and [Fe/H](observed) using equation (4) has a scatter with a range between the extreme outer limits of 0.8 dex. However, the majority of the differences are in the range of ± 0.2 dex (Fig. 8), showing the viability of this Jurscik/Kovacs method to determine remarkably good values of [Fe/H] from the Fourier components.

(4). The rms deviation of calculated minus observed [Fe/H] residuals in the Jurscik/Kovacs formulation using P rather than log P is 0.186 dex for the complete sample in Table 1. If the residuals larger than ([Fe/H] (calculated minus observed) of 0.54 dex are excluded, the rms deviation in ([Fe/H] is 0.108 dex for the Jurscik/Kovacs formulation. Our equation (4) using log P gives an rms deviation of ([Fe/H] (calculated minus observed) of 0.191 dex for the complete sample, and 0.109 dex for the sample with exclusions. Hence, there is no difference between the two formulations.

(5). Equation (4) for the Simon/Kovacs et al. effect is shown to be identical in principal with the log period shift with metallicity when amplitude (Eq. 7), or rise times (Eq. 8), are used as the independent mediating variable. Equation (7) using amplitudes is expected to be useful in estimating mean values of [Fe/H] in large samples of RR Lyrae variables in external galaxies (with, however, the effects of evolution in the calculated [Fe/H] yet remaining as noise). All that is needed are the observed amplitudes and the corrected periods.

(6). The RR Lyrae stars in the two anomalous clusters NGC 6388 and 6441 with high metallicity and with long periods for their amplitudes violate equations (4) and (7), giving spurious calculated values for [Fe/H]. The evident higher luminosities for their horizontal branches are not currently understood, but the reasons in terms of higher helium abundances and therefore higher He core masses are under study (e.g. Catalan & Sweigart; Bono et al.). Such RR Lyrae stars with anomalously long periods for their metallicities evidently are rare because few if any are presently known in the general field.

(7). The evolution toward brighter luminosities of RR Lyrae variables away from the age-zero horizontal branch produces noise in the amplitude/period equation (7). The noise, estimated from the width of the horizontal branch of globular clusters (S90), is similar to the observed scatter in the calculated minus observed residuals in [Fe/H] of 0.2 dex in Fig. 8.

(8). The expected noise due to evolution from equation (4) (or in the Jursciks/Kovacs equation 3) using the observed φ31 values is not present in the currently available globular cluster data, in contradiction to the expectations of the model, for reasons not presently understood.

(9). The problem presents an opportunity to improve the assumptions of the model concerning the precise variations of φ31 at various positions in the instability strip as possible functions of variations of [Fe/H] and luminosity, cluster-to-cluster. It can be anticipated that an understanding will eventually fall out when precision data become available for RR Lyraes in globular clusters of different metallicity.

ACKNOWLEDGEMENTS

Correspondence with Norman Simon at the beginning of the work concerning the meaning of the Fourier components is appreciated. Comments by Geza Kovacs, Horace Smith, Carla Cacciari, Flavio Fusi Pecci and a most helpful anonymous referee were important in improving early drafts. I am grateful to Leona Kershaw for transcribing the manuscript into an acceptable form for electronic submission to the Journal. It is a special pleasure to thank David Sandage for preparing the diagrams for publication.

FIGURE CAPTIONS

Fig. 1. The correlation of rise time and $\varphi_{31}$ from the data in Table 1, flagged for four ranges of [Fe/H]. Skipping-jack crosses are for the metallicity range of [Fe/H] from +0.07 to -1.00 (for a mean of <[Fe/H]>= -0.48); black dots for a metallicity range from -1.01 to -1.40 (mean = -1.29); roman crosses from -1.41 to -1.80 (mean = -1.63); open triangles from -1.81 to -2.60 (mean = -2.08).

Fig. 2. Correlation of $\varphi_{31}$ with absorption-corrected color from the data in Table 1. Symbols are the same as in Fig. 1.

Fig. 3. The $\varphi_{31}$-period correlation for stars in Table 1 for two intervals of [Fe/H] with mean values of -0.35 (8 stars), and -1.63 (16 stars). The diagram is clearly separated into the two regions by [Fe/H]. The actual correlation is a continuum, artificially made discrete by taking mean values in only two intervals of [Fe/H]. The lines are from the adopted three-parameter correlation equation (4) of the text, calculated using the mean [Fe/H] values shown here.

Fig. 4. The correlation of $\varphi_{31}$ with [Fe/H] for two fixed values of period, averaged from the data in Table 1 over the log period intervals of -0.442 to -0.329 (average = -0.367 from 14 stars), and -0.200 to -0.127 (average = -0.169, 15 stars). As in Figure 3, if all stars in Table 1 had been plotted, all regions of this parameter space would be covered in a continuum. The continuum variation with $\varphi_{31}$ is the Fourier-metallicity effect discovered by Simon/Kovacs/Jurcsik/Zsoldos.

Fig. 5. The Preston-like (1959) correlation of metallicity and period for the field stars in Table 1 is in the left panel. The totality of the data in the table is shown. The limit line at small periods (the blue fundamental edge) has the equation log P = -0.452 + 0.033 ([Fe/H])2 derived elsewhere (Sandage 2004), replacing the linear equation of log P = -0.117 [Fe/H] -0.526 in Sandage 1993a (Fig. 10). The right panel shows that the scatter is correlated with $\varphi_{31}$, depicted here as a discrete relation using two mean values of $\varphi_{31}$ and a linear approximation to the quadratic boundary line in the left panel.

Fig. 6. Schematic of the $\varphi_{31}$, log P correlation as a function of metallicity, calculated from equation (4) for six metallicity values. The "period shift" of a representative data point with observed values of $\varphi_{31}$ and log period is shown. The shift is measured relative to the adopted fiducial line for [Fe/H] = 0.00 given by equation (6). The test for the slope of the lines of constant [Fe/H] is for a lack of dependence of the $\Delta$log P values on $\varphi_{31}$, as tested in Figure 7.

Fig. 7. Test for cross talk between the coefficients of the terms in equation (4) by noting any separation of the period-shift values (the ordinate) for small (open circles) and large (skipping-jack) values of $\varphi_{31}$ as a function of [Fe/H](observed).

Fig. 8. The difference between the calculated and observed values of [Fe/H] as a function of log P for all stars in Table 1, separated into large and small values of $\varphi_{31}$ to test for cross talk. The rms metallicity deviation per star is 0.19 dex using all the data. Elimination of the ten largest residuals gives an rms deviation of only 0.11 dex.

TABLE 1

THE RR LYRAE STARS STUDIED BY SIMON AND LEE AND BY SIMON

| Star | [Fe/H] (Layden) | Log P | AV | RT | (B-V)o | (31 | (Log P |
|------|-----------------|-------|------|------|--------|------|--------|
| (1) | (2) | (3) | (4) | (5) | (6) | (7) | (8) |
| AA Aql | -0.58 | -.442 | 1.33 | .14 | (.31) | 1.85 | -.045 |
| RW Tra | +0.07 | -.427 | 0.77 | .21 | (.40) | 2.26 | +.022 |
| HH Pup | -0.69 | -.408 | 1.40 | .13 | (.33) | 1.75 | -.099 |
| V445 Oph | -0.23 | -.401 | 0.90 | .18 | (.44) | 2.29 | +.002 |
| W Crt | -0.50 | -.385 | 1.33 | .13 | .34 | 1.97 | -.078 |
| SW And | -0.38 | -.355 | 0.95 | .19 | .37 | 2.27 | -.048 |
| RV Cap | -1.72 | -.349 | 1.15 | .14 | .30 | 1.36 | -.237 |
| ST Oph | -1.30 | -.346 | 1.31 | .13 | (.33) | 1.64 | -.183 |
| RR Leo | -1.57 | -.344 | 1.38 | .12 | .30 | 1.62 | -.189 |
| V455 Oph | -1.42 | -.343 | 0.90 | .21 | (.27) | 1.71 | -.172 |
| VX Her | -1.51 | -.342 | 1.35 | .16 | ---- | 1.21 | -.274 |
| SW Aqr | -1.24 | -.338 | 1.31 | .14 | (.25) | 1.58 | -.203 |
| CP Aqr | -0.90 | -.334 | 1.31 | .14 | .34 | 1.98 | -.127 |
| DN Pav | -1.37 | -.329 | 1.39 | .12 | .32 | 1.47 | -.235 |
| DX Del | -0.56 | -.326 | 0.75 | .18 | (.34) | 2.33 | -.065 |
| V440 Sgr | -1.42 | -.321 | 1.29 | .12 | (.29) | 1.69 | -.198 |
| ST Leo | -1.29 | -.321 | 1.24 | .13 | .33 | 1.77 | -.182 |
| SV Hya | -1.70 | -.320 | 1.32 | .20 | (.25) | 1.45 | -.248 |
| RY Col | -1.11 | -.320 | 0.70 | (.20) | .39 | 1.89 | -.159 |
| BB Pup | -0.57 | -.318 | 1.06 | .14 | (.31) | 2.16 | -.107 |
| BR Aqr. | -0.90 | -.317 | 1.19 | .16 | .37 | 2.00 | -.140 |
| AV Ser | -1.20 | -.312 | 1.14 | .18 | .39 | 1.79 | -.187 |
| SS For | -1.35 | -.305 | 1.36 | .17 | .32 | 1.82 | -.183 |
| TY Aps | -1.21 | -.300 | 1.09 | .20 | (.29) | 2.19 | -.119 |
| RZ Cet | -1.50 | -.292 | 0.98 | .18 | .35 | 1.73 | -.219 |
| V499 Cen | -1.56 | -.283 | 1.22 | .13 | .33 | 1.65 | -.244 |
| AN Ser | -0.04 | -.282 | 1.05 | .17 | .41 | 2.54 | -.066 |
| RY Psc | -1.39 | -.276 | 0.84 | .18 | .34 | 1.91 | -.199 |
| VY Lib | -1.32 | -.273 | 1.04 | .16 | (.36) | 1.98 | -.188 |
| RW Gru | -2.00 | -.260 | 1.06 | .13 | .34 | 1.69 | -.259 |
| RR Cet | -1.52 | -.257 | 0.98 | .16 | .37 | 1.95 | -.210 |
| V452 Oph | -1.72 | -.254 | 1.02 | .16 | (.39) | 1.77 | -.249 |
| RV Oct | -1.34 | -.243 | 1.16 | .15 | (.36) | 2.06 | -.202 |
| TZ Aqr | -1.24 | -.243 | 0.86 | .16 | .40 | 2.18 | -.178 |
| WY Ant | -1.66 | -.241 | 0.93 | (.24) | .31 | 1.76 | -.264 |
| V341 Aql | -1.37 | -.238 | 1.27 | .16 | (.28) | 0.02 | ---- |
| RX Eri | -1.30 | -.231 | 0.90 | .18 | .39 | 2.12 | -.202 |
| V413   | -2.33 | -.196 | 1.19 | .16 | .32   | 1.68 | -.325 |
| BH Peg | -1.38 | -.193 | 0.64 | .21 | (.40) | 2.42 | -.180 |
| UY Boo | -2.49 | -.187 | 1.18 | .14 | .37   | 1.61 | -.348 |
| X Ari  | -2.40 | -.186 | 0.99 | .14 | (.34) | 1.86 | -.299 |
| AV Vir | -1.32 | -.183 | 0.79 | .19 | .42   | 2.40 | -.194 |
| SU Dra | -1.74 | -.180 | 1.03 | .16 | .33   | 2.18 | -.241 |
| TV Leo | -1.97 | -.172 | 1.29 | .22 | .36   | 1.93 | -.299 |
| BO Aqr | -1.80 | -.159 | 1.13 | .18 | .38   | 2.16 | -.266 |
| TY Pav | -2.31 | -.148 | 0.92 | .18 | (.37) | 2.22 | -.265 |
| VY Ser | -1.82 | -.146 | 0.73 | .23 | .40   | 2.30 | -.251 |
| X Crt  | -1.75 | -.135 | 0.71 | .19 | .40   | 2.60 | -.201 |
| IU Car | -1.85 | -.132 | 1.01 | .17 | (.34) | 1.92 | -.341 |
| AT Ser | -2.05 | -.127 | 0.92 | .20 | .34   | 2.26 | -.278 |

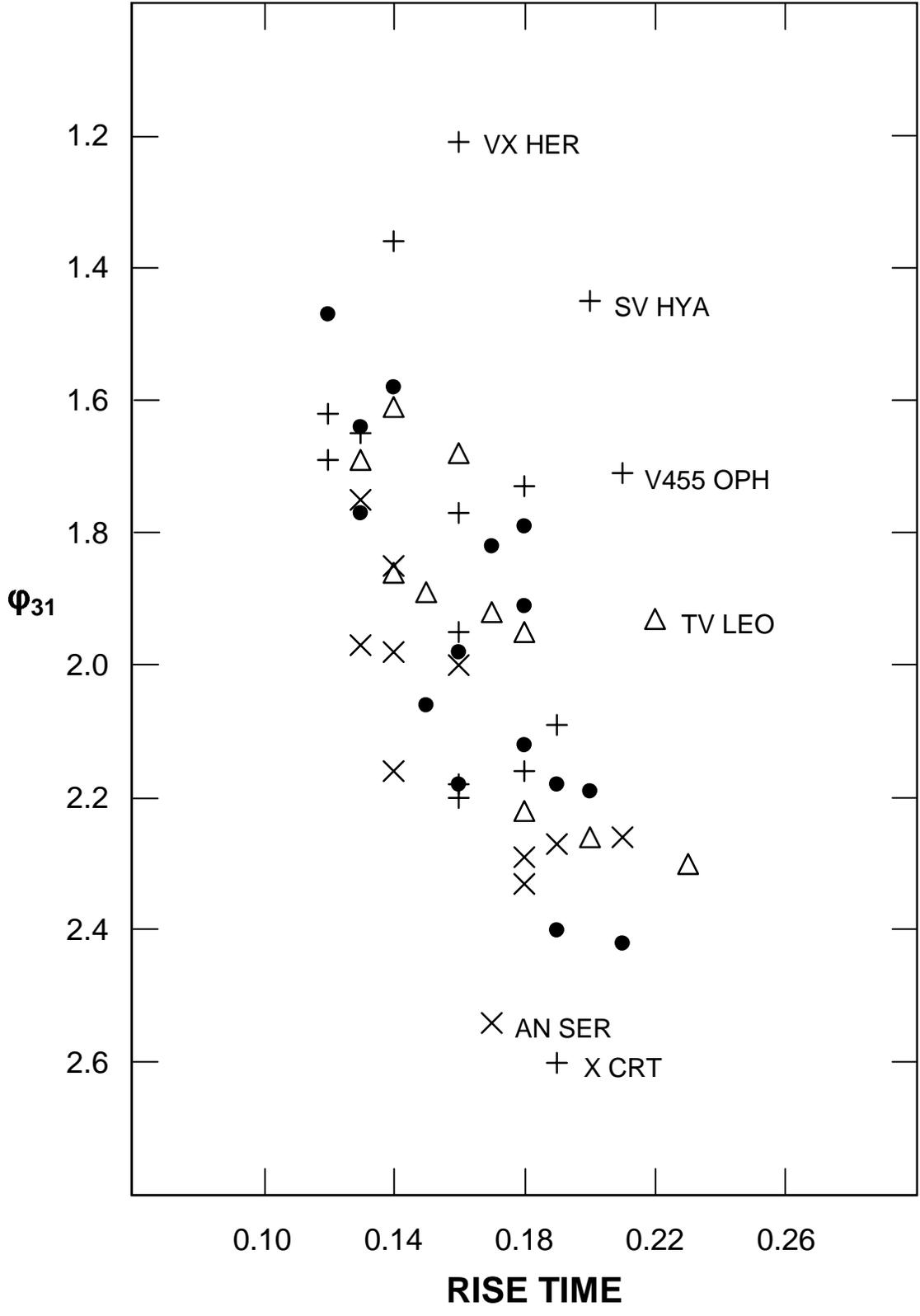

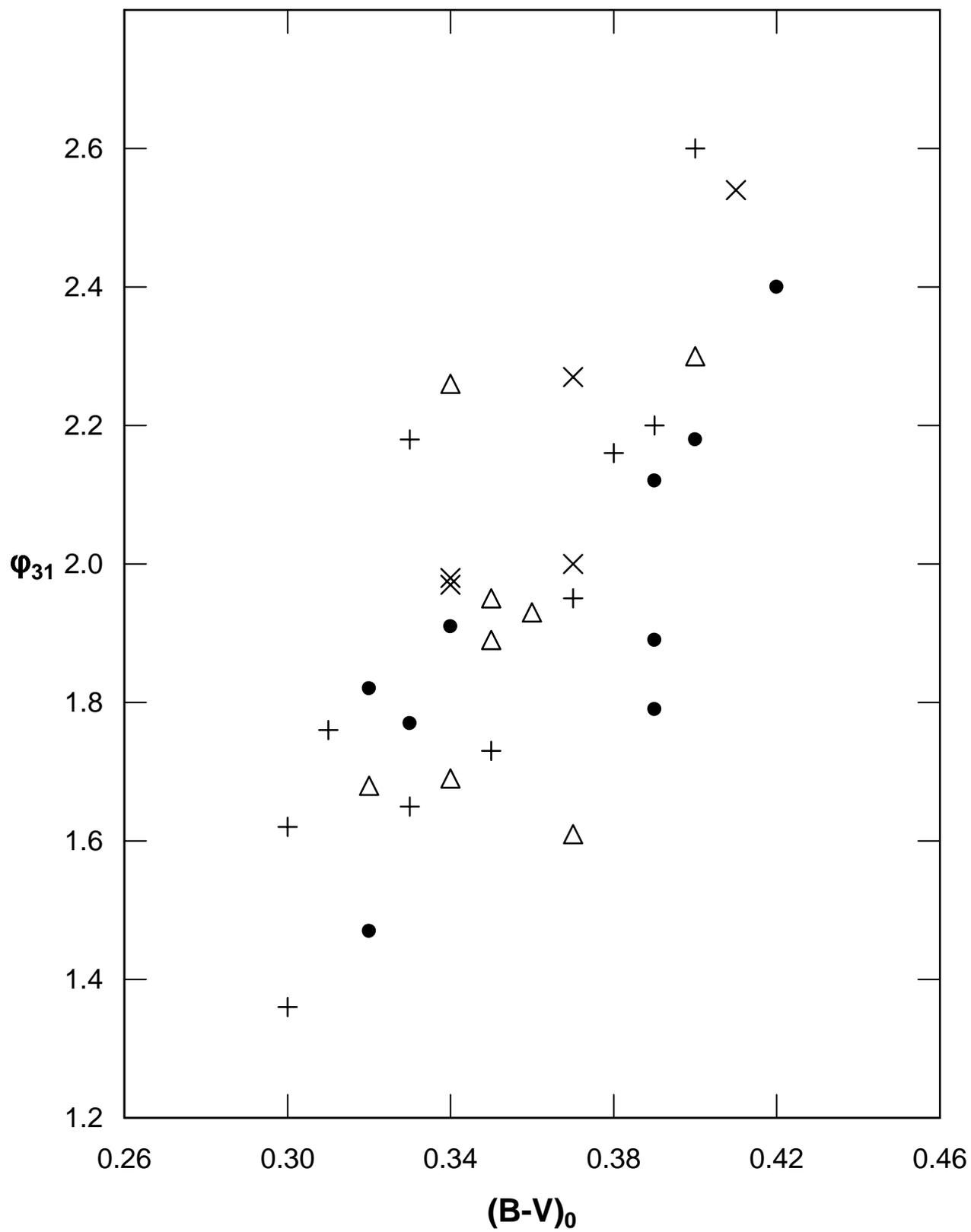

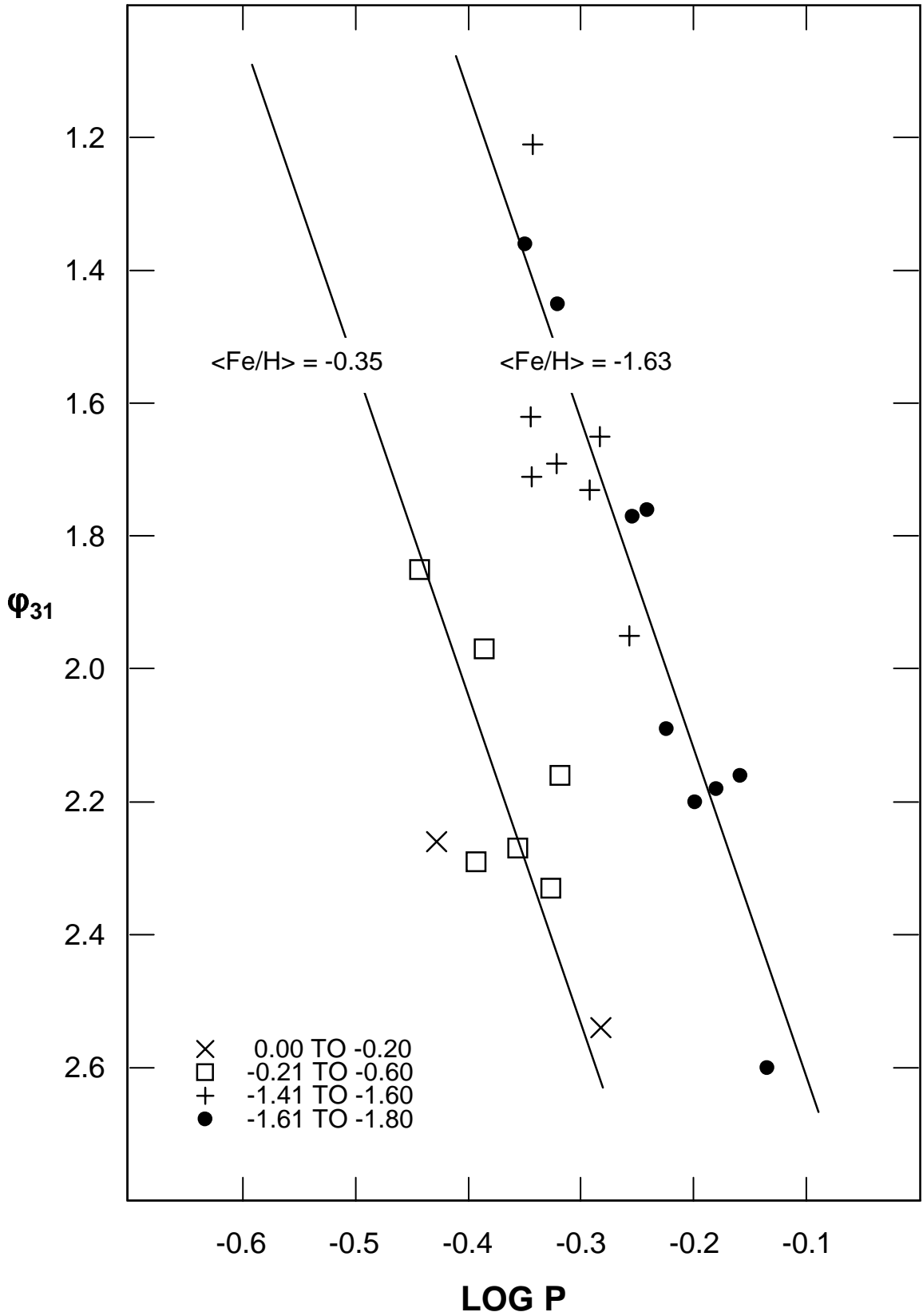

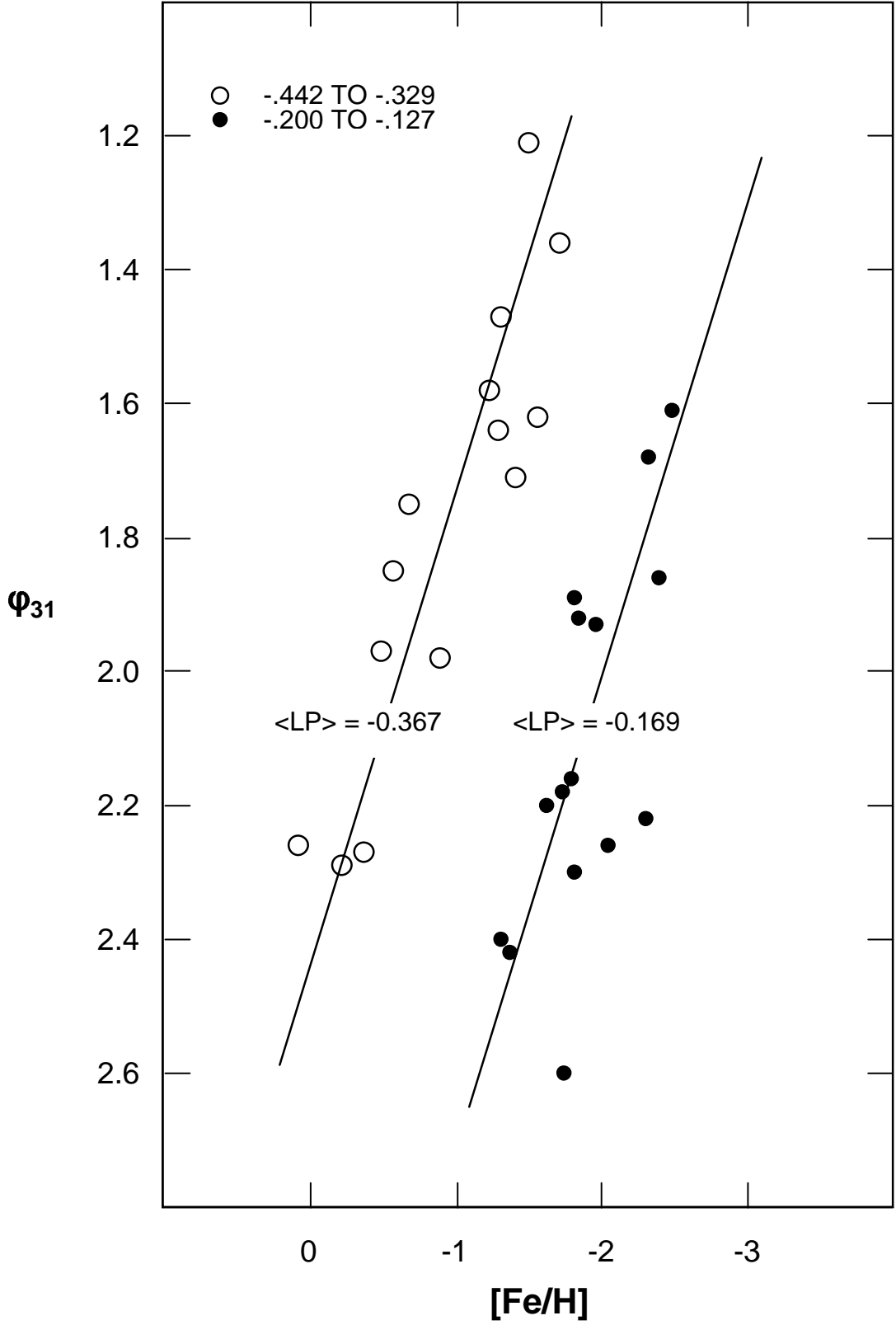

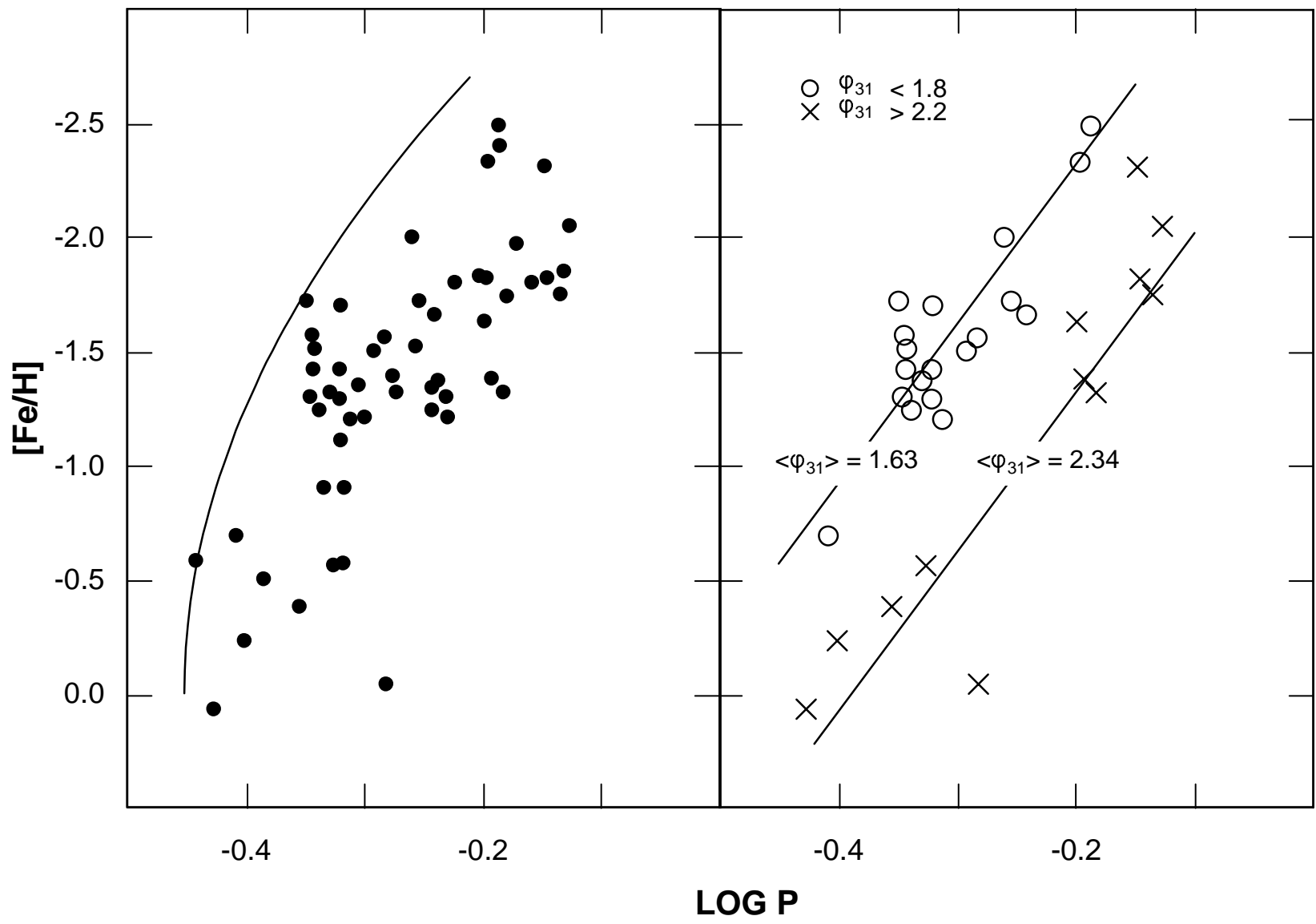

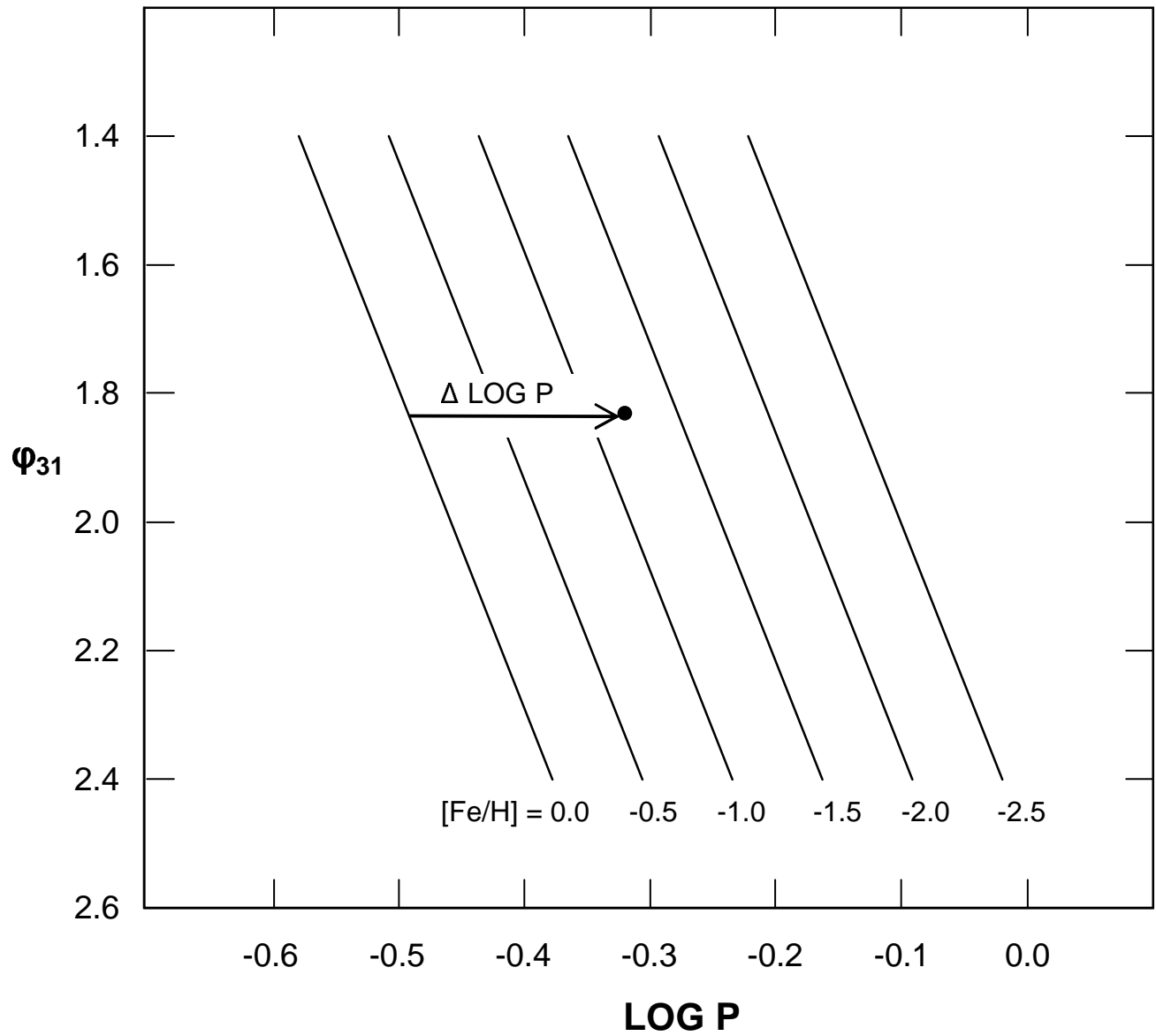

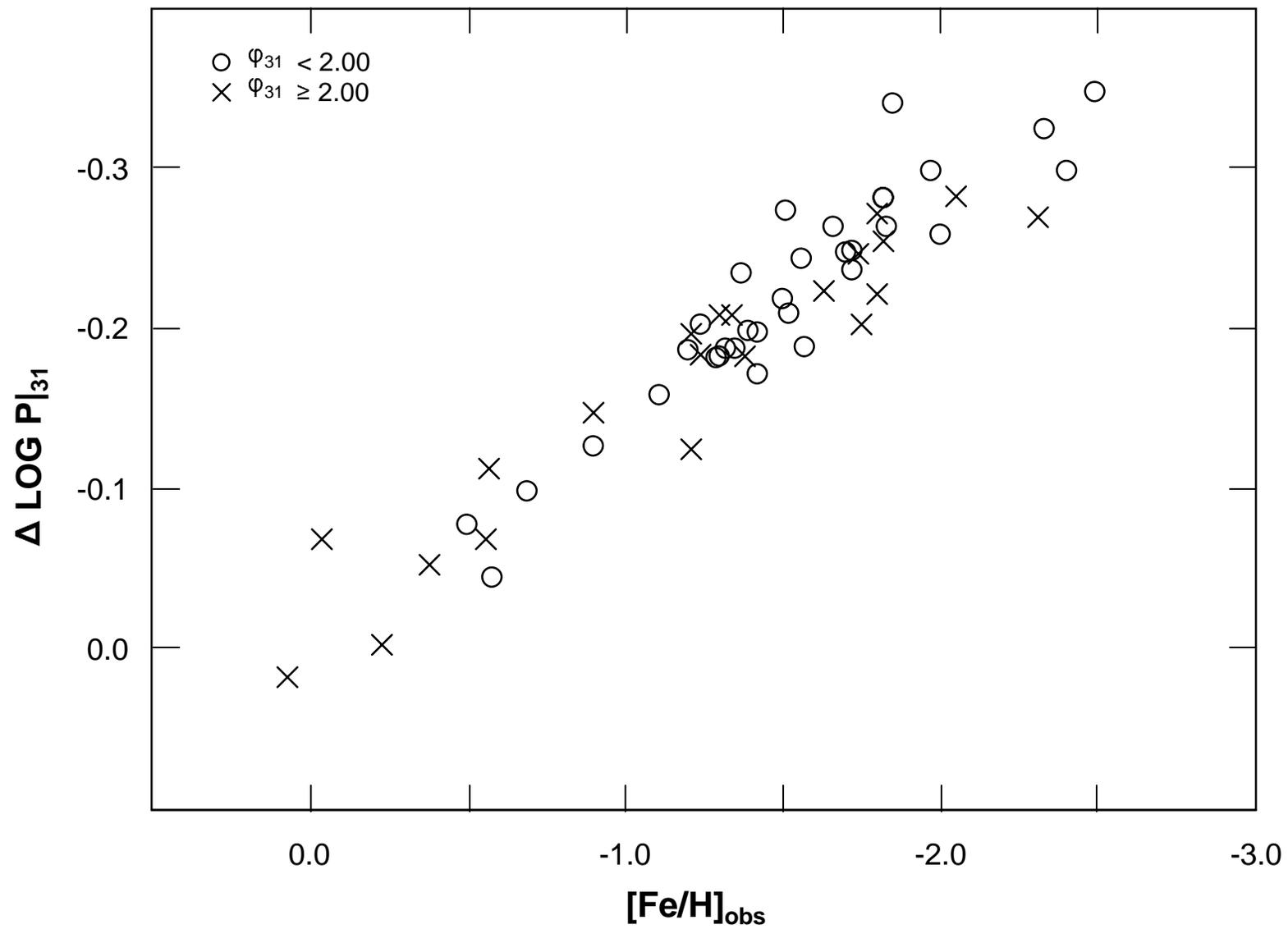

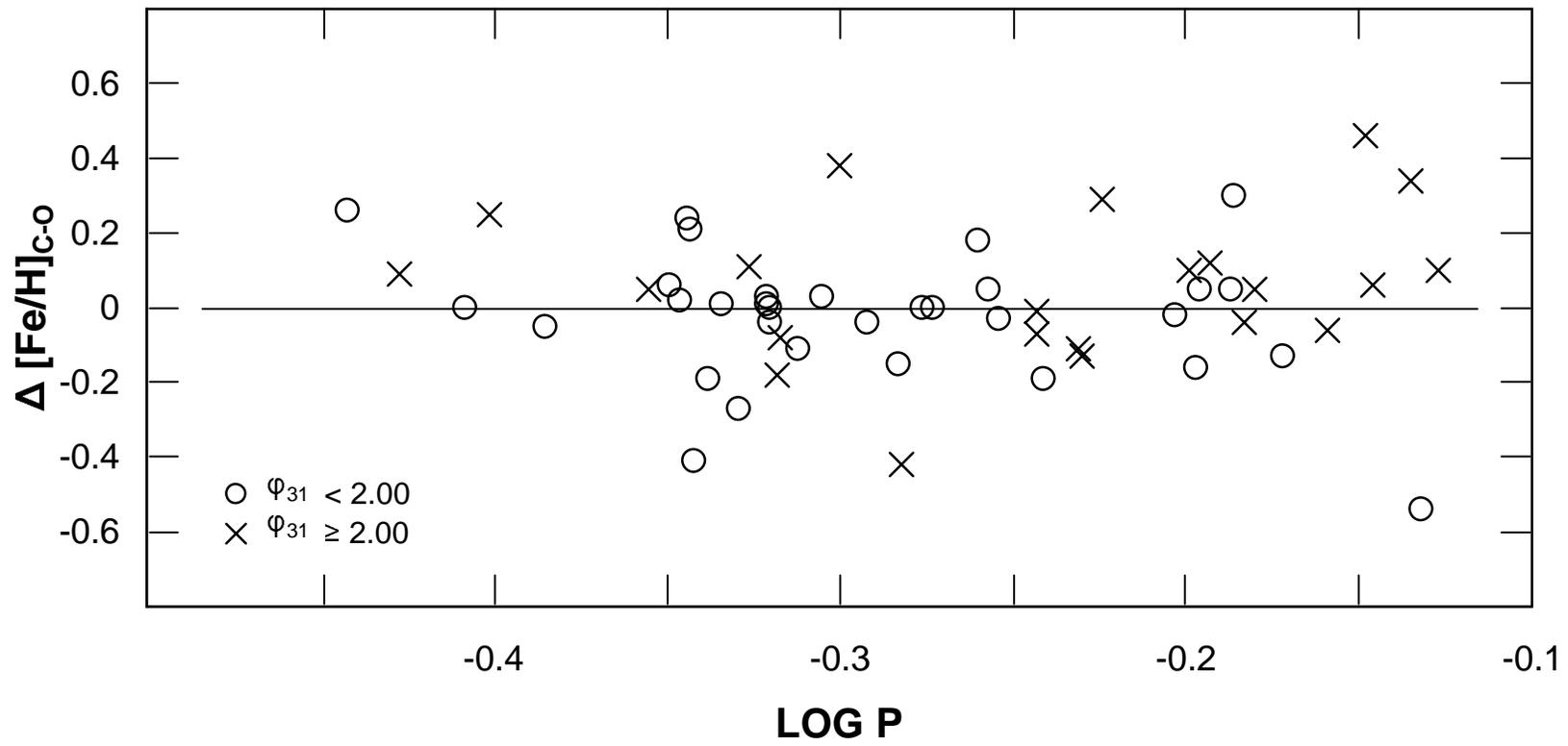